\documentclass[review]{elsarticle}

\usepackage{hyperref}

\journal{To be decided}

\biboptions{authoryear}

\usepackage{float}
\usepackage{graphicx}
\usepackage{textcomp}
\usepackage{natbib}
\usepackage{amsmath}
\usepackage{subfig}
\usepackage{lineno}
\usepackage{rotating}
\usepackage{color}
\usepackage{bm}
\usepackage{verbatim}
\usepackage{multirow}

\newcommand{\comments}[1]{}

\newcommand{\cd}[1]{\texttt{#1}}

\begin{document}

\begin{frontmatter}

\title{Development of a Flexible Coupling Framework for Coastal Inundation Studies.}



\author[address1,address2]{Saeed Moghimi \corref{mycorrespondingauthor}}
\cortext[mycorrespondingauthor]{Corresponding author
(Saeed Moghimi; Email: Saeed.Moghimi@noaa.gov)}

\author[address3]{Andre Van der Westhuysen \corref{_andre}}
\author[address3]{Ali Abdolali \corref{_andre}}
\author[address1]{Edward Myers \corref{_andre}}
\author[address1]{Sergey Vinogradov \corref{_andre}}
\author[address3]{Zaizhong Ma  \corref{_andre}}
\author[address4]{Fei Liu  \corref{_andre}}
\author[address3]{Avichal Mehra \corref{_andre}}
\author[address5]{Nicole Kurkowski \corref{_andre}}

\address[address1]{​NOAA Coast Survey Development Laboratory, National Ocean Service, Silver Spring, USA.} 
\address[address2]{University Corporation for Atmospheric Research, Boulder, USA. }
\address[address3]{​NOAA Center for Weather and Climate Prediction, College Park, USA.} 
\address[address4]{​NOAA Earth System Research Laboratory, Boulder, USA.} 
\address[address5]{​NOAA Office of Science and Technology Integration, Silver Spring, USA.} 

\cortext[_andre]{\textbf{Co-authors contribution:}\\ 
1. \textbf{Saeed Moghimi}: Responsible for development of the ADCIRC, WW3-data
and HWRF-data Caps and developing the coupled application. Writing the body of the manuscript and plotting figures.
\\2. \textbf{Andre Van der Westhuysen}: Responsible for unstructured WW3 cap
development.
Inclusion of the WW3 cap in the coupled application and running full coupled cases; manuscript preparation and analysis of the results.
\\3. \textbf{Ali Abdolali}: Responsible for WW3 cap development and running
stand alone wave model, plotting wind and wave related parts of manuscripts, manuscript preparation and analysis of the results.
\\4. \textbf{Edward Myers}: Scientific support, discussion and project
management; \\5. \textbf{Sergey Vinogradov}: Scientific support and discussions.
\\6. \textbf{Zaizhong Ma}: Responsible for wind field modeling using HWRF.
\\7. \textbf{Fei Liu}: Support for ESMF/NUOPC cap development.
\\8. \textbf{Avichal Mehra}: Scientific support and project management.
\\9. \textbf{Nicole Kurkowski}: Scientific support and project management.
}

\begin{abstract}
To enable flexible model coupling in coastal inundation studies, a coupling
framework based on \cd{ESMF/NUOPC} technology under a common modeling framework
called the NOAA Environmental Modeling System (\cd{NEMS}) was developed. The
framework is essentially a `software wrapper around atmospheric, wave and storm 
surge models that enables its components communicate seamlessly, and efficiently 
run in massively parallel environments.   
We  implemented  the coupled application  including  \cd{ADCIRC} and
unstructured \cd{WWAVEWATCHIII} caps as well as \cd{NUOPC} compliant caps to read Hurricane Weather Research and 
Forecasting Model (\cd{HWRF}) generated forcing fields.    
We validated the coupled application for a laboratory test and a full scale
inundation case of the Hurricane Ike, 2008, on a high resolution mesh covering the whole US
Atlantic coast.
We showed that how nonlinear interaction between surface waves and total water level
 results in significant enhancements and progression of the inundation
 and wave action into land in and around the hurricane landfall region. We also
 presented that how the maximum wave setup and maximum surge regions may happen
 at the various time and locations depending on the storm track and geographical
 properties of the landfall area.

\end{abstract}

\begin{keyword}
wave-current interaction\sep tidal inlet \sep river plume \sep wind waves \sep three-dimensional circulation  \sep wave breaking 
\end{keyword}

\end{frontmatter}


\section{Introduction}
\label{sec:intro}

To establish a coastal flooding modeling system, several model
components based on the target geographical region need to be coupled.  To
accurately simulate the total water level in a tropical hurricane land-falling
inundation study, a dynamically coupled system of numerical models including
storm surge, surface waves, inland river flooding and numerical weather
prediction are necessary. On top of that based on the geographical location
other model components may need to be employed. For instance, to setup an
efficient coastal flooding prediction system for Alaska region, inclusion of a
sea-ice model is essential. 


In recent years, Earth System Models were proven to be invaluable tools that
enabled us to better understand and more accurately predict our environment. 
Each system includes a coupled applications that consists of several model
components to represent relevant physical processes.  The model components
are expected to interact with each other similar to what takes place in nature.

There are several Earth System Model software flavors that enable model
components to communicate by importing and exporting data
\citep{jacob2005m, valcke2012coupling, hill2004architecture}.
The Earth System Modelling Framework (\cd{ESMF}) has been utilized to develop
several earth system coupled applications worldwide
\citep[e.g.][]{warner2008development,moghimi2012three,lemmen2017modular}. To
increase \cd{ESMF} interoperability, the National Unified Operational Prediction Capability 
(\cd{NUOPC}) consortium developed a layer consisting of a set of generic components
 \citep{theurich2016earth}. 
\cd{NUOPC} layer is a software wrap around \cd{ESMF} and was developed
collaboratively by several research and operational centers. The primary
objectives behind \cd{NUOPC} design are to be reusable, extensible and
portable framework for ESM coupling.


The NOAA Environmental Modeling System (\cd{NEMS}) is a coupled modeling
infrastructure designed to address increasing needs for prediction of the earth
environment at a range of time scales.  \cd{NEMS} includes several external
model components that have a primary source code outside NOAA. Therefore, NOAA
only needs to maintain and develop the coupling interfaces (so-called model
caps) of the given modeling component. In turn, the \cd{NEMS} ecosystem allows
connecting various combinations of model components into a number of different
coupled model applications to address specific environmental phenomena at
specific time scales.         

The present research goals are to develop a flexible and generic coupling
between ADvanced CIRCulation model \citep[\cd{ADCIRC};][]{luettich1992adcirc}
and \cd{WAVEWATCH III} \citep[\cd{WW3;}][]{tolman2009user} via their respective
\cd{NUOPC} caps, and to provide an infrastructure to make future development and inclusion of various model
components, such as river and inland flooding coupling, seamlessly possible. The
current development of the \cd{NUOPC} caps provides the possibility to perform
dynamical coupling of \cd{ADCIRC} and the unstructured version of \cd{WW3}, as
well as various atmospheric models \cd{ATM}. The cap developed for \cd{ADCIRC}
is capable of importing atmospheric forcing and surface wave fields, and exporting water
surface elevation and current velocity to the connected model components.
Conversely, the cap developed for unstructured \cd{WW3} imports atmospheric
forcing, water levels and current, and exports the wave radiation stresses
required to force \cd{ADCIRC}.         

The first application of this new coupled system is the so-called Named Storm
Event Model (\cd{NSEM}), a high-fidelity model for hindcasting coastal
inundation and total water level. It is being developed to meet the requirements
of the NOAA's Consumer Option for an Alternative System to Allocate Losses
(COASTAL) Act of 2012. This modeling system includes \cd{ADCIRC} as the 
hydrodynamic component, \cd{WW3} as the wave model, the Hurricane
Weather Research and Forecasting Model (\cd{HWRF}) as the atmospheric component
\citep{tallapragada2014evaluation}, and in future the National Water Model
(\cd{NWM}) as the inland  hydrological component \citep{gochis2013wrf}, see Fig.
\ref{fig:cpl-app}. 

The structure of this paper is as follows. First, we describe the envisioned
design of the \cd{NEMS ADCIRC-WW3} coupled application and the methodology. 
Then a detailed description of the \cd{ADCIRC} cap implementation and available
coupling options is given. This is followed by a similar description of the
\cd{WW3} cap. Subsequently, we present the results of the coupled system. Finally 
we present verification of the coupled \cd{ADCIRC-WW3} application for the
laboratory flume case of \cite{boer1996Surfzone}, as well as a full scale
storm surge inundation event during Hurricane Ike, 2008 in the US Gulf of
Mexico.

\section{Structure of the coupled application}
\label{sec:coupled}

A typical \cd{NUOPC} application includes a number of generic components that
provide an interface to the underlying \cd{ESMF} infrastructure for generating
and operating a coupled application in a fairly straightforward and seamless
manner. The generic components are defined as follows:    
\begin{itemize}
  \item A \textit{Driver} manages all the components to initialize, run,
  finalize and keep track of time for exchanging information among model
  components.    
  \item \textit{Connectors} are used to execute field matching, grid remapping
  and data redistribution among model components.  
  \item	A \textit{Model} (cap) wraps each model component code (e.g. \cd{ADCIRC}
  and \cd{WW3}) to provide  a generic interface and standard metadata suitable
  to be plugged into the \textit{Driver}, and form a multi-model coupled
  application. 
  \item An optional \textit{Mediator} wraps custom coupling code to calculate
  quantities which  includes data from several model components or requires
  operations such as time averaging. 
\end{itemize} 

The system includes methods and utilities for time management, error handling,
high performance inputs/outputs (\cd{I/O}), grid remapping and field
interpolation.  Since \cd{NUOPC} is a layer around \cd{ESMF} library, function
calls to both \cd{NUOPC} and \cd{ESMF} are possible and sometimes are
necessary.       

In this research, we developed a \cd{NUOPC} application that includes a driver,
three \cd{NUOPC} enabled model components and four connectors. The components
are not allowed to directly access each other's data. The only way the data
moves in or out of a component is via instances of an \cd{ESMF\_state} class.
The state is a container that wraps native data and also includes a metadata to
let the other components know about name, coordinates and decomposition of the
actual packed data. 

\begin{figure}
    \centering
    \hspace*{-12mm}
    \includegraphics[width=0.45\textwidth]{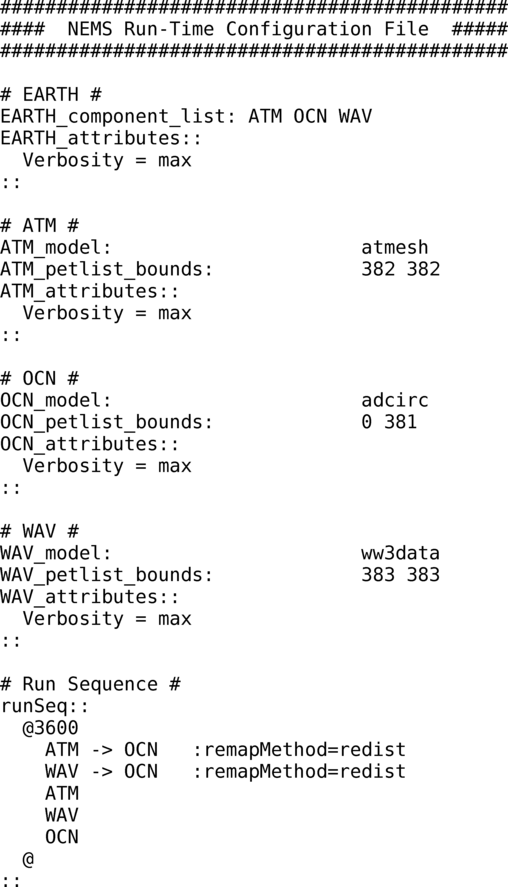}
    \caption{Coupled application configuration file. \\  \textit{All model configurations and
    results are pre-decisional and for official use only}.}
    \label{fig:cpl-conf}
 \end{figure}

The driver component accesses \cd{ADCIRC}, \cd{WW3} and \cd{ATMesh} model
components via their \cd{SetServices()} methods. It reads basic information for
how to initialize and run the model components from a configuration text file 
(Fig. \ref{fig:cpl-conf}).  The configuration file contains information about
name of the model components, number of processes to be associated to each model
component, the coupling time intervals, and the order of data exchange among the
components.  The driver also initializes the number of connectors by providing
the name of the sending and receiving model components.  Therefore, for a
dynamical two-way coupling between two model components, two connectors are
required.           

The connector component initializes at the run time by matching the list of
available import and export fields advertised by the model components. The
connector establishes the connection based on matched import and export fields.
The connector also has access to the domain decomposition and computational
domain discretization of the connected model components. It will generate a
remapping and necessary weight matrices for interpolation of the fields among
model components at the initialization phase. In other words, the connector
receives exported data in the form of an \cd{ESMF\_state} in the native grid or
mesh from the exported model component and passes it to the importing model
components in its own native grid or mesh definition. This remapping facility
allows the coupled system the freedom to use different meshes for, say, the
circulation and wave modeling components, and/or different domain
decompositions. This is useful in cases where the wave model component requires
a different mesh optimization to resolve its distinct physics, or more
computational cores for load balancing.               

\begin{figure}
    \centering
    \hspace*{-12mm}
    \includegraphics[width=0.75\textwidth]{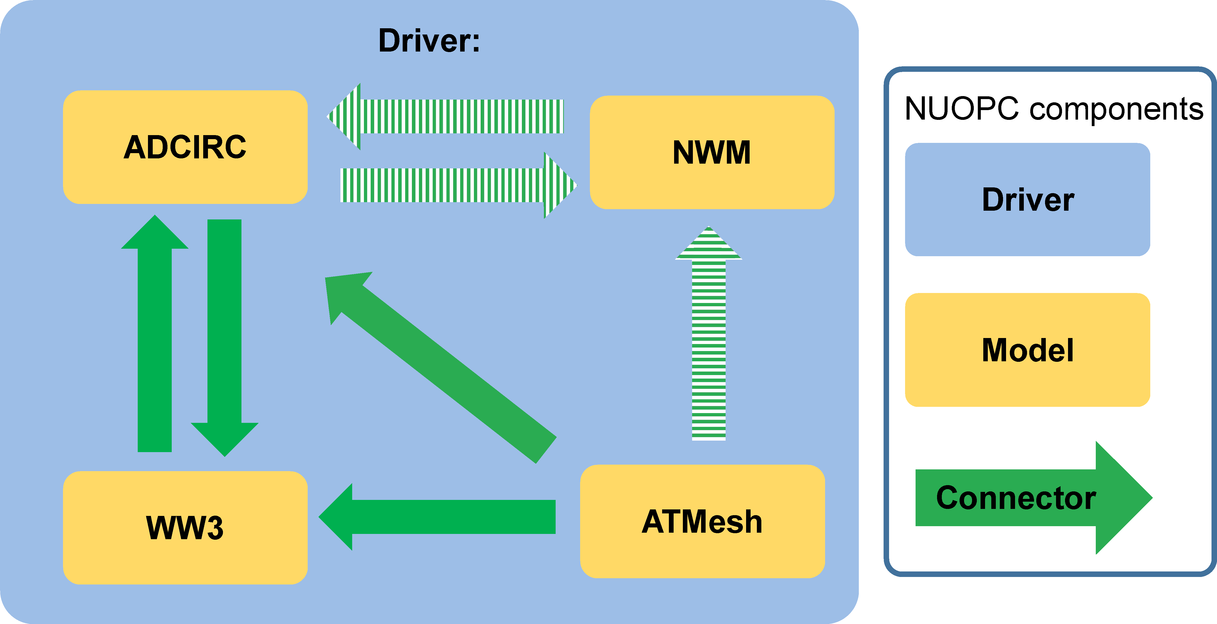}
    \caption{Design of the coupled application for coastal flooding inundation
    studies (\cd{NSEM}). \\ \textit{All model configurations and
    results are pre-decisional and for official use only}. }
    \label{fig:cpl-app}
 \end{figure}

The \cd{ATMesh} cap was developed as a placeholder interface for a full live
atmospheric model, which was not included in our NSEM application due to scope
limitations. This so-called data cap reads weather prediction outputs (from a
NetCDF data file), initialize required \cd{NUOPC/ESMF} objects and provide
requested data and information to \cd{ADCIRC} and \cd{WW3} caps via the
\cd{NOUPC/ESMF} backbone. The \cd{NWM} hydrological model
component and its associated connectors are not yet implemented (Fig. \ref{fig:cpl-app}).       

\section{ADCIRC model}
\label{sec:adcirc}
The ADvanced CIRCulation model (\cd{ADCIRC}) is a finite element hydrodynamic
community model originally developed by \cite{luettich1992adcirc}. \cd{ADCIRC} 
is undergoing continuous development by groups of scientists and engineers. Its
natural finite element unstructured mesh capability, and several modules
specifically addressing various aspects of the coastal flooding and tropical
cyclone forcing, make it one of the best tools available for coastal inundation
studies. \cd{ADCIRC} operates in either two-dimensional depth-integrated (2D)
depth-averaged (barotropic) and three-dimensional (baroclinic) modes.  In the 2D mode, it
solves equations for both water surface elevation and the depth-averaged
velocity fields.  For more details about \cd{ADCIRC} governing equations,
numerical methods and wave forcing implementation please see
\citep{luettich1992adcirc,dietrich2011modeling}.            

\cd{ADCIRC} is written in modular \cd{FORTRAN} and supports parallel execution
on massive supercomputers using \cd{MPI} architecture.  The code structure is
partitioned in three distinct initializing, running and finalizing phases ready
for the ESMF coupling.  The model initializes by a call to \cd{ADCIRC\_Init()}
which also receives a \cd{MPI}  communicator from the driver. The subroutine
reads necessary input files for constructing the computational mesh including
nodes location and connectivity.  It also builds a local and global nodal map to
reference which nodes reside on which \cd{MPI} process, and to identify their
global relationships.  It reads input information to constrain the model such as
bathymetry, meteorological forcing, and freshwater inflow and open boundary
conditions.  As a part of the initialization, ADCIRC also checks and connects to
all requested output files that will be used as containers to fill in the model
results.             

\cd{ADCIRC} enters the run phase by a call to \cd{ADCIRC\_Run()} subroutine,
which also receives an argument for the number of time steps (\cd{NTIME\_STP})
for that specific run request. The start time and end time of the simulation is
determined during the initialization phase. The model run takes place via a time
loop in which, at every time step, a single call to the \cd{TIMESTEP()}
subroutine occurs. All the computational steps for applying forcing and boundary
conditions to produce the final results are being performed in this subroutine.         
The \cd{ADCIRC} concludes its run by a call to \cd{ADCIRC\_Final()} subroutine
where some of the final post-processing and check for \cd{MPI} finalizing are
performed.    

\subsection{\cd{ADCIRC} coupling interface (cap)}
\label{sec:adcirc_cap}

The \cd{ADCIRC} \cd{NUOPC} cap performs the coupling in all the three phases:
initialize, run and finalize. In the development of the \cd{NUOPC} cap for
\cd{ADCIRC}, extreme care and attention were paid to minimize changes to the
original \cd{ADCIRC} code. At the initialization of the \cd{NUOPC} application,
a global \cd{MPI} communicator is created by \cd{ESMF} infrastructure and a
dedicated set of processes passed to \cd{ADCIRC} via a \cd{MPI} communicator
based on the number of processes requested for \cd{ADCIRC} in the configuration
file. At the initialization, \cd{ADCIRC} cap also gets connected to available
import and export field matches accepted by the communicators.

After information exchange among the model components, the
\cd{ModelAdvance()} subroutine of the \cd{ADCIRC} cap calls the
\cd{ADCIRC\_Run()} subroutine to perform the next run interval. Tab.
\ref{tab:import_export} shows the list of the exported and imported fields
currently accepted by the \cd{ADCIRC cap}.  The naming conventions of these
variables are defined in the \cd{NUOPC} field dictionary to allow
interoperability with other \cd{NUOPC} components. We modified and tested \cd{ADCIRC}
preprocessing and main model code to accommodate various coupling arrangements. 
The \cd{NWS} input parameters in \cd{fort.15} input file
are described in Tab.  \ref{tab:nws} \citep{moghimi2019development}.          

\begin{table}
 \begin{center}
   \caption{Exported and imported fields in coupled system. \\ \textit{All model configurations and
    results are pre-decisional and for official use only}.}
   \label{tab:import_export}
   \begin{tabular}{l l l l l }                      \hline
    Data field   &  Units         & Variable    & Exported by  & Imported by \\ 
    \hline
    Eastward sea water velocity	& $m s^{-1}$  & \cd{UU2} & \cd{ADCIRC}& \cd{WW3}\\ 
    Northward sea water velocity& $m s^{-1}$  & \cd{VV2} & \cd{ADCIRC}& \cd{WW3}\\
    Sea surface height above mean sea level	& $m$ 	 & \cd{ETA2} & \cd{ADCIRC}& \cd{WW3}\\
    \hline 
    Eastward radiation stress	 & $m^2s^{-2}(N m^{-2}/\rho)$ &
    \cd{ADCIRC$\_$SXX} & \cd{WW3} &  \cd{ADCIRC} \\
    Northward radiation stress	 & $m^2s^{-2}(N m^{-2}/\rho)$ &
    \cd{ADCIRC$\_$SYY}& \cd{WW3} &  \cd{ADCIRC} \\
    Cross radiation stress	     & $m^2s^{-2}(N m^{-2}/\rho)$ &
    \cd{ADCIRC$\_$SXY}& \cd{WW3} &  \cd{ADCIRC} \\
    \hline 
    Air pressure at sea level    & $m H_2O$	   & \cd{PRN2}& \cd{ATMesh}& \cd{ADCIRC}\\ 
    Eastward wind at 10m height  & $m s^{-1}$  & \cd{WVNX2}& \cd{ATMesh}& \cd{ADCIRC}, \cd{WW3}\\ 
    Northward wind at 10m height & $m s^{-1}$  & \cd{WVNY2}& \cd{ATMesh} & \cd{ADCIRC}, \cd{WW3}\\ 
    \hline
   \end{tabular}
 \end{center}
\end{table}

\begin{table}
 \begin{center}
   \caption{New implemented and tested \cd{ADCIRC} options. \\ \textit{All model configurations and
    results are pre-decisional and for official use only}.}
   \label{tab:nws}
   \begin{tabular}{l l l}                      \hline
    \cd{NWS} parameter &  Meteorological forcing     & Wave forcing   \\ \hline 
    17            &  ATM$^*$                    & None         \\
    517           &  ATM$^*$                    & WAV$^{**}$   \\
    500           &  None                       & WAV$^{**}$   \\
    519           & Best Track (Holland Model)  & WAV$^{**}$   \\
    520           & Best Track (Generalized Asymmetric Holland Model) & 
    WAV$^{**}$ 
    \\
    \hline
  \end{tabular}\\
  \end{center}
    $^* ~$ Any \cd{NUOPC} enebaled numerical weather prediction model providing
    required data fields e.g. \cd{ATMesh} cap.\\
    $^{**}$ Any \cd{NUOPC} enebaled wave model model providing required data
    fields e.g. \cd{WW3} cap.\\
\end{table}

\section{WAVEWATCH III}

\cd{WAVEWATCH III (WW3)} \citep{tolman2009user} is a third-generation spectral wave model
that solves the wave action balance equation that accounts for the growth,
propagation, non-linear interaction and dissipation of wind waves in the ocean
by:     
\begin{eqnarray}\label{actbal}
\frac{\partial N}{\partial t} + \nabla_x \cdot (\mathbf{c}_g + \mathbf{U})N +
\frac{\partial}{\partial k}\dot{k}N + \frac{\partial}{\partial \theta}
\dot{\theta} N = \frac{S}{\sigma}  
\end{eqnarray}
where $N(k,\theta)$ is the wave action density spectrum, related to the wave
energy density spectrum $F(k,\theta)$ where $N(k,\theta) =
F(k,\theta)/\sigma$. Here $k$ is the wavenumber and $\dot{k}$ its propagation
speed due to depth- or current-induced Doppler shifting, $\theta$ is the wave
direction and $\dot\theta$ its propagation speed due to depth- or current
refraction, $\sigma$ is the wave frequency, $\mathbf{c}_g$ is the group
velocity, and $\mathbf{U}$ the depth-averaged current velocity. On the
right-hand side, $S$  represents the sum of source terms, including wave growth,
nonlinear interaction and dissipation. \cd{WW3} was originally developed on a
regular grid for global operational wave forecasting, with 2-way nesting for
regional applications. More recently, it has been extended to curvilinear grids
for Arctic applications \citep{rogers2018forecasting}, as well as unstructured meshes for
high-resolution coastal application \citep{ardhuin2013development}. Most
recently, its traditional `card deck' MPI parallel implementation
\citep{TOLMAN200235} has been supplemented with a more conventional domain-decomposition
 approach using ParaMETIS \cite{roland2012fully} equipped with an optional
 implicit equation solver. Along with these improvements in numeric, source terms suitable for nearshore application have been added, including
depth-induced breaking, reflection, three-wave non-linear interaction, and
wave-ice interaction, amongst others.                    

The \cd{WW3} code is written in modular \cd{FORTRAN}, similar to \cd{ADCIRC},
and is broken up into a collection of sub-programs which are run in sequence to
carry out a simulation. The most important of these are \cd{ww3\_grid}, which
compiles the computational mesh, and physics and numerics settings into
\cd{mod\_def.ww3}, a binary resource file, \cd{ww3\_prep}, which preprocesses all
forcing files, \cd{ww3\_multi}, the multi-grid core wave model, and
\cd{ww3\_ounf} and \cd{ww3\_ounp}, which are NetCDF postprocessing routines. To
comply with the \cd{ESMF} protocol, the core wave model \cd{ww3\_multi} has
been broken up into \cd{w3init}, \cd{w3wave}, and \cd{w3final} to perform the main
steps of model initialization, model advancing, and model finalization,
respectively \citep{campbell2013coupling}. During the initialization step with
\cd{w3init}, the configuration of the computational mesh file, including node indices,
geographical location, and the mesh connectivity is read from the binary
resource file \cd{mod\_def.ww3}. Using the domain decomposition from the
\cd{PDLIB} library, a local and global nodal map is built that references which
nodes reside on which MPI process, and how they are related globally. During the
model advance step \cd{w3wave}, forcing fields such as water depth, wind
velocity and currents as well as boundary conditions are updated, followed by
the solution of the wave action equation equation by means of fractional
stepping. Output is written to a set of binary output files for later
postprocessing (using \cd{ww3\_ounf} and \cd{ww3\_ounp}). Model finalization is
completed by calling \cd{w3final}.                       

\subsection{WAVEWATCH III \cd{NUOPC} cap}
The \cd{WW3} \cd{NUOPC} cap carries out the coupling in the three phases
(initialization, advancing and finalizing) described above. A regular grid
version of this cap was developed by \cite{campbell2013coupling} for global and
regional-scale \cd{NUOPC} applications. In the present work, this cap was
extended to support unstructured meshes and domain decomposition for
high-resolution coastal modeling. During the initialization of the \cd{NUOPC}
application, import and export meshes are defined based on the \cd{PDLIB}
decomposition, a global \cd{MPI} communicator is created by the \cd{ESMF}
infrastructure, and a dedicated set of processes are passes to \cd{WW3} via a
\cd{MPI} communicator based on the number of processes requested for \cd{WW3} in
the configuration file (Fig. \ref{fig:cpl-conf}). During this initialization step, the \cd{WW3} cap is
also connected to available import and export field matches accepted by the
communicators.

After the information exchange among the model components, the
\cd{ModelAdvance()} subroutine of the \cd{WW3} cap calls the \cd{w3wave}
subroutine to perform the next run interval. Tab. \ref{tab:import_export} shows
the list of the exported and imported fields accepted by the \cd{WW3} cap for
the current application. It is noted that a larger set of import and export
variables is supported by the \cd{WW3} cap in general, including surface
roughness variables, Stokes drift and bed roughness, used in other NUOPC
applications featuring wind waves. For more details, see
\cite{campbell2013coupling}.

\subsection{Wave-induced stresses}
Breaking waves transfer their momentum to ocean currents. Mathematically, this
forcing is expressed in the circulation model as the divergence in the radiation
stresses, as described in some detail below. From spectral wave models such as
\cd{WW3}, the radiation stress vectors can be evaluated from the computed wave
energy density as follows \citep{tolman2009user}:      
\begin{eqnarray}\label{SXXSYYSXY}
S_{XX}&=\rho_w g \int\int(n-0.5 +n\cos^2(\theta))F(k,\theta)d\theta dt \nonumber \\
S_{XY}&=\rho_w g \int\int n\sin(\theta)\cos(\theta)F(k,\theta)d\theta dt \nonumber \\
S_{YY}&=\rho_w g \int\int(n-0.5 +n\sin^2\theta)F(k,\theta)d\theta dt 
\end{eqnarray}
where $\rho_w$ is the water density, $g$ is the gravitational acceleration, the
directional wave energy density spectrum $F(k,\theta) = \sigma N(k,\theta)$, $d$
is the water depth, and $n$ is the ratio of the wave group $c_g$ to wave phase
speed $c_p$ for a given depth and frequency, given by:     
$n=\frac{1}{2}+\frac{kd}{\sinh(2kd)}$.

In order to account for the impact of the waves on the mean circulation, the
spatial gradient of wave radiation stress per unit area $\tau_{s, waves}$ are
calculated as:
\begin{eqnarray}\label{FXFY}
\tau_{sx,waves} = -\left( \frac{\partial S_{XX}}{\partial x}+\frac{\partial S_{XY}}{\partial y} \right)\nonumber \\
\tau_{sy,waves} = -\left( \frac{\partial S_{YY}}{\partial y}+\frac{\partial S_{XY}}{\partial x} \right) 
\end{eqnarray}
and incorporated as additional surface stresses alongside wind stresses and
bottom stresses into \cd{ADCIRC}'s Generalized Wave Continuity Equation and
vertically-integrated momentum equations, following \cite{dietrich2011modeling}.

\section{Model setup}
\label{sec:model-setup}

We utilized the existing Hurricane Surge On-demand Forecast System
unstructured triangular mesh as the base of the
computational domain for the coupled model setup. The HSOFS mesh covers the entire Gulf of Mexico
and extends into the Atlantic Ocean to the approximate longitude of 65$\circ$W, allowing for appropriate
generation of storm surge from atmospheric effects over a large region. The HSOFS 
mesh has 1.8 M nodes and covers the shallow coastal regions up to a 
topographic height of 10~m above local mean sea level with the mesh resolution of approximately 
250~m (See Sec. \ref{sec:hurr-ike}.

    \hspace*{-12mm}


\section{Results}
\label{sec:results}

The importance of dynamical coupling of surge and surface waves on the spatial extent
of the inundation and active wave action area were investigated for Hurricane Ike, 2008.

Hurricane Ike was a powerful tropical cyclone that swept through portions of the
Greater Antilles and Northern America in September 2008, wreaking havoc on
infrastructure and agriculture, particularly in Cuba and Texas. The ninth
tropical storm, fifth hurricane, and third major hurricane of the 2008 Atlantic
hurricane season, Ike developed from a tropical wave west of Cape Verde on
September 1 and strengthened to a peak intensity as a Category 4 hurricane over
the open waters of the central Atlantic on September 4 as it tracked westward.
Several fluctuations in strength occurred before Ike made landfall on eastern
Cuba on September 8. The hurricane weakened prior to continuing into the Gulf of
Mexico, but increased its intensity by the time of its final landfall on
Galveston, Texas on September 13.  

The wave-surge coupled application (hearafter Fully coupled) and stand
alone models (hearafter ``Stand alone'') were forced with an identical HWRF
meteorological forcing (See Sec. \ref{Sec:Atmosphericforcing}). 
As a reference, the \cd{ADCIRC} model results forced with
tidal boundary condition is also presented (heresafter ``Only tide'').

\begin{figure}
    \hspace*{10mm}
    \textbf{a)}
    \includegraphics[width=0.9\textwidth]{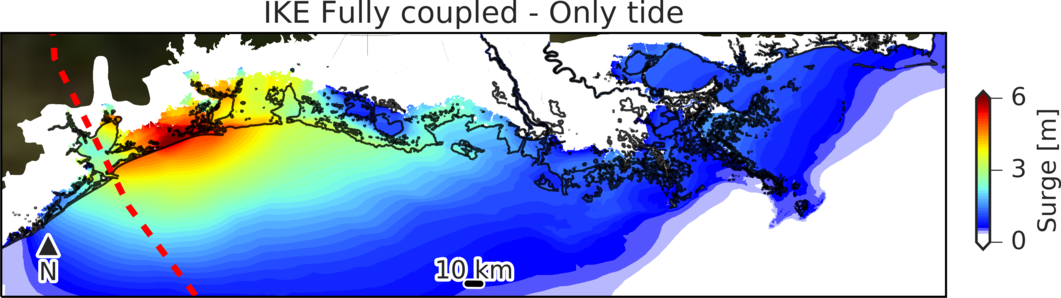}
    \\
    \hspace*{10mm}
    \textbf{b)}
    \includegraphics[width=0.9\textwidth]{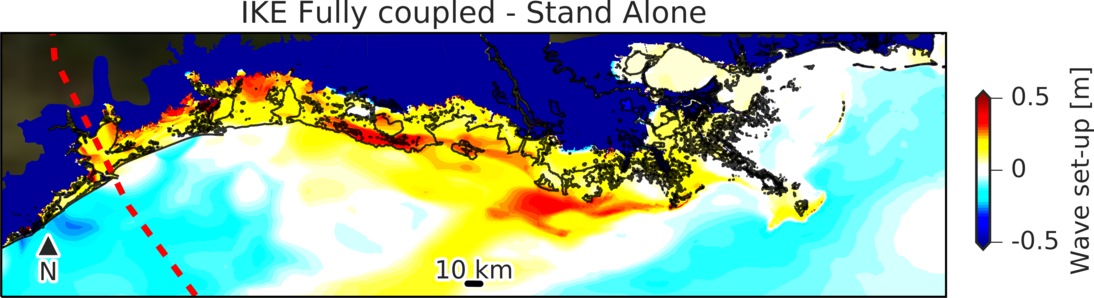}
    \caption{Total surge level computed by subtracting the tide only run from
    fully coupled case which includes both atmospheric and wave coupling
    contributions (a). The maximum wave contribution in
    total water level computed by subtracting stand alone case from fully
    coupled (b).
    Red line represents the Hurricane Ike best track.
    Black contour line represents the shoreline, and the areas beyond the black
    contour line are the inundated regions. \\  \textit{All model configurations
    and  results are pre-decisional and for official use only}.} 
    \label{fig:wav-effect}
\end{figure}

The map of the maximum surge level during the whole simulation for the
Fully coupled case is shown in Fig. \ref{fig:wav-effect}a. The maximum surge 
level is calculated by subtracting tidal water level from Fully coupled results.
The results reveal that the most severe inundation during Ike, 2008 with more than 6~m above the
maximum tide level was took place on the east side of the hurricane track in the
region between the Galveston Bay, Tx and the Sabine Lake, Tx. 


The maximum wave contribution to total water level is calculated by subtracting 
Stand alone water level from Fully coupled results (Fig. \ref{fig:wav-effect}b).
It is shown that some of the wave induced inundation (wave setup) occurs at the edge of the
maximum surge where the atmospheric wind setup and negative pressure are at
their maximum strength. This is significant because it shows how wave induced momentum
released from breaking waves increases the total water level which in turn causes the wave active breaking
region to advance further into the land and release wave action on the structures
further landwards. 

\begin{figure}
    \centering
   \includegraphics[width=0.9\textwidth]{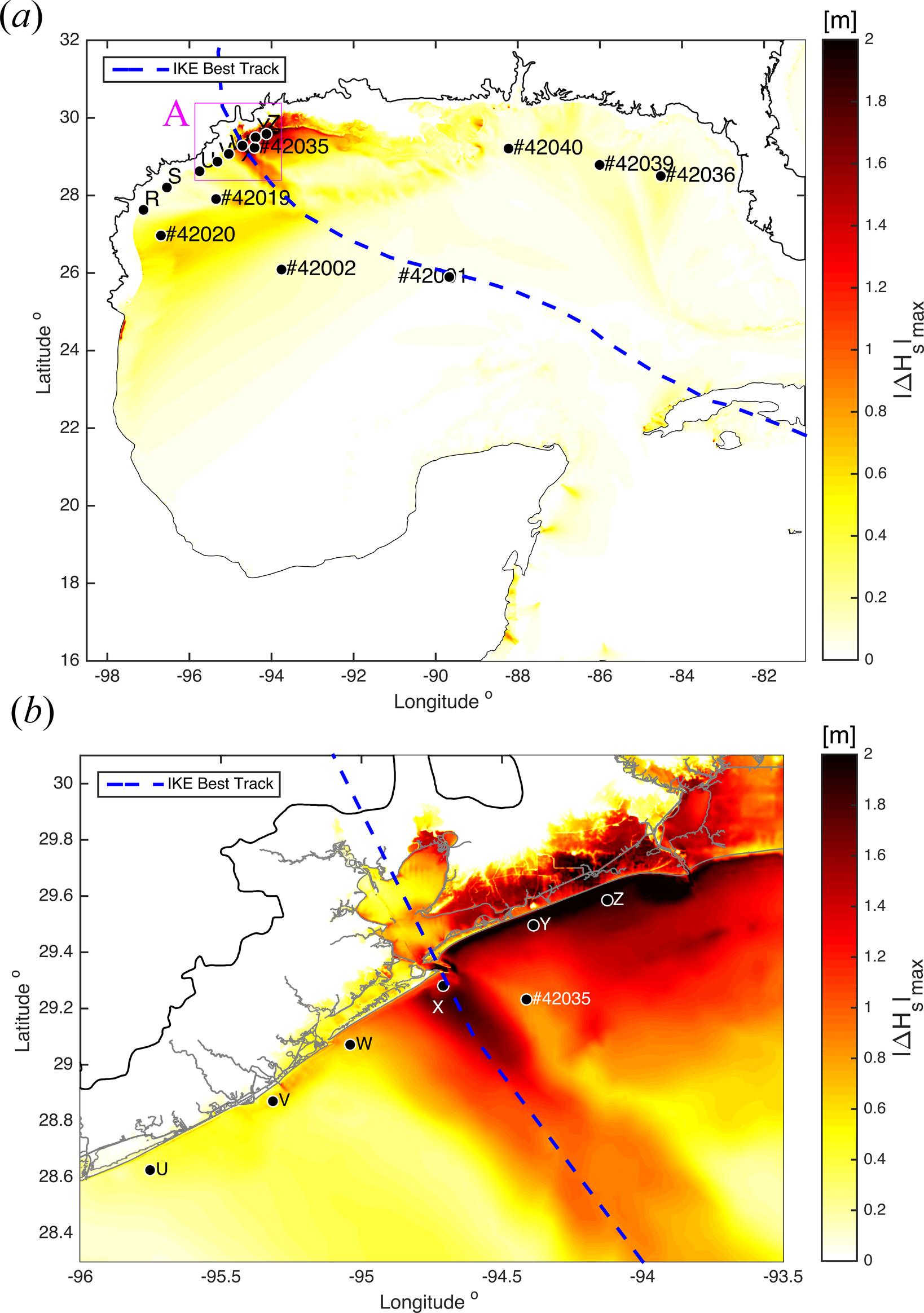}
\caption{Wave model sensitivity to dynamic exchange between wave and surge
models in term of the spatial distribution of the absolute difference between
significant wave height $H_s$, extracted from the fully coupled Wave-Surge and
Stand Alone WW3 models; Gulf of Mexico (a) landfall region(b). \\
\textit{All model configurations and  results are pre-decisional and for
official use only}.}      
    \label{fig:DIF_CO_ST}
 \end{figure}

Wave height significantly enhanced due to the dynamical coupling of the surge and
wave components. Fig. \ref{fig:DIF_CO_ST}a, b represents the maximum wave
height difference during the whole storm between Fully coupled and Stand 
alone (which includes no tidal or surge/inundation effects)
cases. Wave height increases more than 2~m along the track as well as
in the east of the Galveston Bay at the coastal and landfall region. The
comparison at the 6 quick deployed wave gauges shows significant contribution of
the surge on the eastern side of hurricane track in nearshore region (Fig.
\ref{fig:WW3AND}. 

It should be noted that wave setup contribution seen in Fig. 
\ref{fig:wav-effect}b at the Mississippi and Atchafalaya rivers delta
region occurred hours before the actual landfall therefore the maximum surge and 
wave setup in this region did not happen at the same time. However,
it also points to the possibility of experiencing large swells and rip-current
events hours before hurricane makes its actual landfall.

\begin{figure}
    \centering
    \hspace*{-12mm}
    \includegraphics[width=0.6\textwidth]{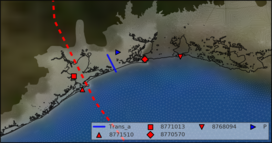}
    \caption{Red markers are the locations of the tidal gauges. The legend shows
    the stations ID numbers. Red dashed line is the Hurricane Ike best track.
    The blue line and blue triangle are transect and time serie plot shown in
    Fig. \ref{fig:trans_a} and time series of the test point is presented
    in Fig. \ref{fig:ts_test_point} respectivly.  \\
    \textit{All model configurations and results are pre-decisional and for official use only}. }
    \label{fig:ts-map}
\end{figure}

\begin{figure}
    \centering
    \hspace*{-12mm}
    \includegraphics[width=1\textwidth]{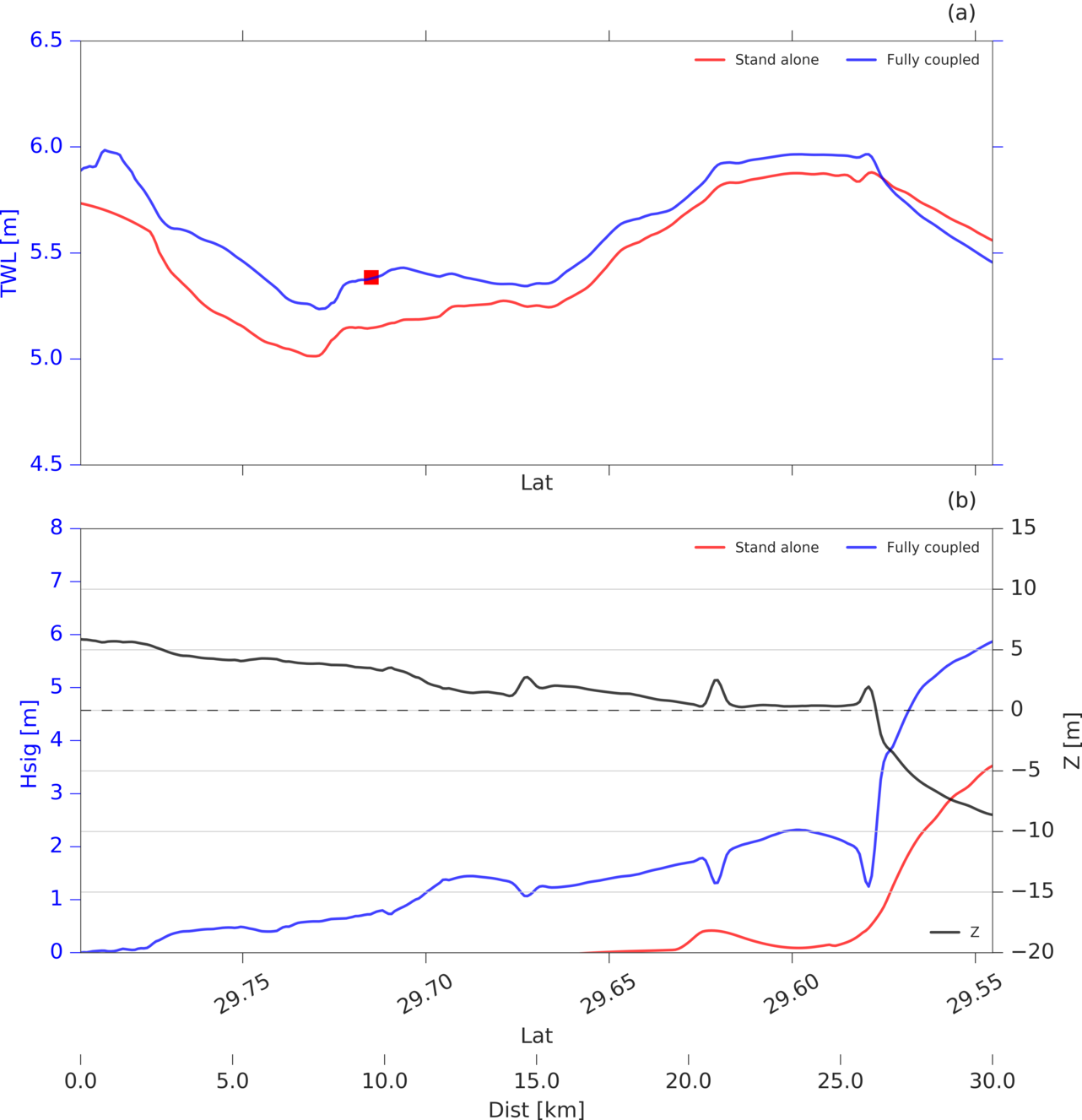}
    \caption{Comparison of the results for the transect(blue line in Fig.
    \ref{fig:ts-map}) of the total water level (a) and wave height
    (b) for Fully coupled and Stand alone cases are presented. The
    bold black line represents topobathy level refrenced to the mean sea
    level (MSL). \\ \textit{All model configurations and
    results are pre-decisional and for official use only}.}
    \label{fig:trans_a}
\end{figure}

To further analyze this mechanism, we plot changes of the total water level (TWL)
and wave height for the Fully coupled and Stand alone cases at a transect  
shown in Fig. \ref{fig:trans_a} (The location of the transect is shown in Fig.
\ref{fig:ts-map}). We also shown the topobathy values along the transect in a
positive upward vertical coordinate system where Z=0~m located at the local mean
sea level (black line in \ref{fig:trans_a}b). We plotted High Water Mark
observations in the 1~km radius from the transect with red squares.
The TWL and wave height line plots start from the land (kilometer 0) in which ground level is
almost 6~m above the local mean sea level (latitude $\sim$ 29.80~N) and continues
towards the ocean to 8~m water depth (kilometer 30) in the ocean side (latitude $\sim$ 29.55~N).
Fully coupled solution for the total water level continuously show enhancement
over the Stand alone results (Fig.\ref{fig:trans_a}a). Around the shoreline
where the Z$\sim$0~m and farther off-shore, the Stand alone model show greater
total water level in comparison with the Stand alone solution.
This directly relates to wave height and its dissipation shown in Fig.
\ref{fig:trans_a}b. 

We see that Stand alone model (forced without water surface elevation from surge
component) shows wave height of almost zero right after waves cross the
shoreline (landwards of the Z=0~m; Kilometer$\sim$24). On the other hand, the
wave height from the coupled model show significant wave height of $\sim$ 2.2~m in the same
region (landwards of the Z=0~m; Kilometer$\sim$24). This pattern is visible
for the wave height evolution even more landwards. This mechanism in
which greaer total water level ($\sim$0.5~m) leads to the potential for more
active local wave generation and propagation and therefore greater
wave dissipation and release wave action which re-ignite enhancement of total
water level in inundated region is presented here.

\begin{figure}
    \centering
    \hspace*{-12mm}
    \includegraphics[width=0.9\textwidth]{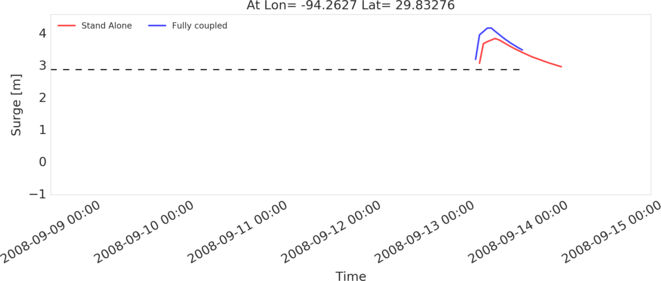}
    \caption{Time series of the total water level at the test point (locations
    shown in Fig. \ref{fig:ts-map}). Black dahsed-line is the ground level.  \\
    \textit{All model configurations and results are pre-decisional and for official use only}.}
    \label{fig:ts_test_point}
\end{figure}

We also looked at a timeseries of water level at  P shown as a blue triangle in 
the Fig. \ref{fig:ts-map}. This point is located very close to the maximum 
spatial extent of the inundated region (Fig.
\ref{fig:ts_test_point}) which is helpful to examine time variation of the total
water level for Full coupled and Stand alone cases. The ground level is
also plotted by a black dashed-line which shows that P is not wet before the
storm as it is located $\sim$3~m above the local mean sea level. During the land
fall both Fully coupled and Stand alone cases produced total water level that
inundated the area. Full coupled shows innudation of $\sim$1~m above the ground
level which is $\sim$0.5~m above the water level resulted by Stand alone case at
the time of the peak of the storm.

%
%
%

    \hspace*{-12mm}

\section{Model system validation}
\label{sec:validation}
We verified the coupled \cd{ADCIRC-WW3} application in a step-by-step manner. In the first
step, we verified all the \cd{ESMF}  intermediate exchange \cd{ESMF$\_$state} fields before sending
and after receiving by the other model component. Then we performed a basic
verification using a small setup to make sure the coupled \cd{ADCIRC-WW3} application runs 
smoothly. Finally, we switched to the full scale inundation test case for
Hurricane Ike, 2008.

\subsection{Laboratory Case (Boers 1996)}
\begin{figure}
    \centering
    \hspace*{-12mm}
    \includegraphics[width=0.7\textwidth]{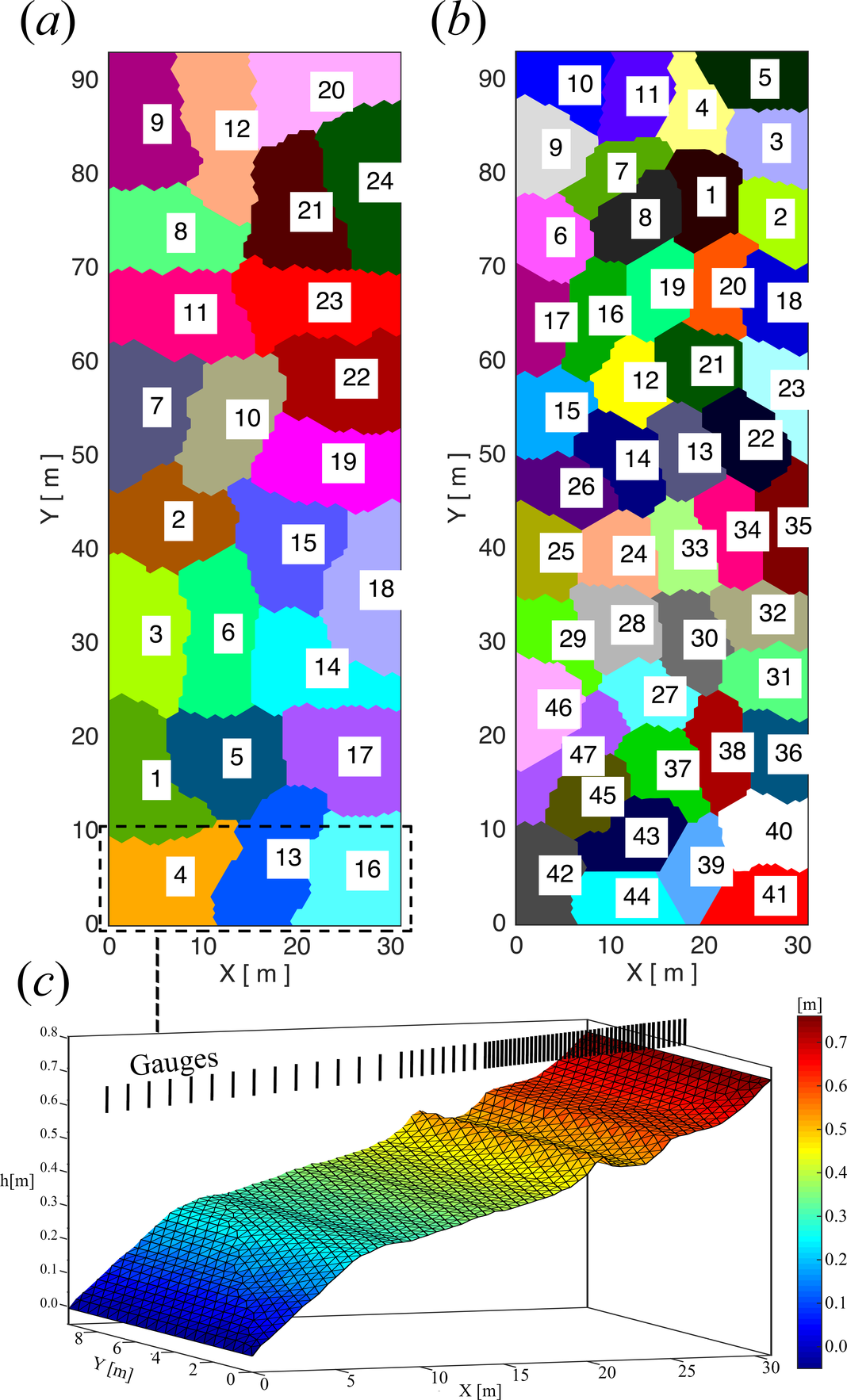}
    \caption{Schematic view of parallelization via domain decomposition
    algorithm in (a) WW3 with 24 subdomains; and (b) ADCIRC with 47 subdomains
    for\cite{boer1996Surfzone} experiment. (c) The numerical domain with
    unstructured triangulated mesh.  \\  \textit{All model configurations and   results are
    pre-decisional and for official use only}.}  
    \label{fig:flume_domain}
 \end{figure}

\begin{figure}
    \centering
    \hspace*{-12mm}
    \includegraphics[width=0.7\textwidth]{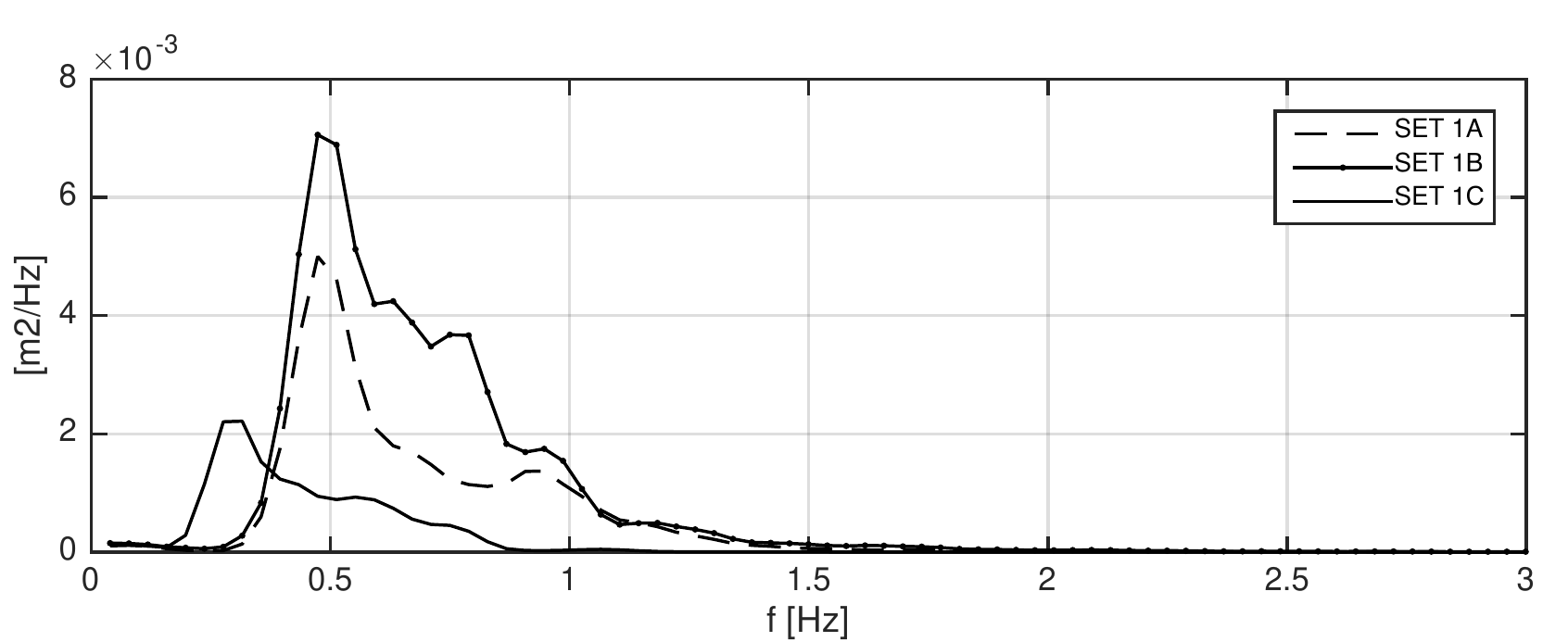}
    \caption{Energy spectra of the three different wave conditions which are 
    summarized in Tab. \ref{tab:boers}. \\ \textit{All model configurations and
    results are pre-decisional and for official use only}.}
    \label{fig:boer_spectra}
\end{figure}

To validate the \cd{ADCIRC-WW3} coupled system, we first
compared the coupled system results against the laboratory flume
experiment of \cite{boer1996Surfzone}. This experiment was carried out at Delft
University of Technology in Spring 1993 to investigate the interaction 
between wind waves and the mean circulation via wave dissipation
and radiation stress transfer. The geometry of the flume is shown in Fig.
\ref{fig:flume_domain}c. The bathymetry represents an immovable (concrete)
profile of a typical barred beach along the western Dutch coastline. The
\cite{boer1996Surfzone} experiment features three test cases, 1A, 1B and 1C, 
in which 1A and 1B represent violently-breaking, locally-generated wind waves, 
and 1C represents a
mildly-breaking swell. Tab. \ref{tab:boers} shows the wave height and peak
period parameters of the imposed wave conditions at the wave maker, while Fig.
\ref{fig:boer_spectra} shows their corresponsing energy spectra.                  

    \hspace*{-12mm}

Fig. \ref{fig:flume_domain}a,b show the domain decompositions of 
the \cd{WW3} and \cd{ADCIRC} model components, respectively. In both cases the
decomposition was generated using METIS \citep{karypis2011metis}, but recall
that in \cd{NUOPC/ESMF} they do not necessarily need to match - for \cd{WW3} the
decomposition was done for 24 cores, while for \cd{ADCIRC} it was done for 47 cores. The wave component of the coupled
\cd{ADCIRC-WW3} was forced with the observed spectra at the upstream wave maker,
and the water level in the flume is initially set to rest. The \cd{WW3}
component is configured with a time step of 1 s. The relevant wave physics
processes are depth-induced breaking and three-wave nonlinear interaction, using
the source term formulations of, respectively, Battjes and Janssen (1978), with
$\gamma_{BJ} = 0.80$ and $\alpha = 1$, and \cite{eldeberky1996spectral}.
The \cd{ADCIRC} model component was run at a time step of 1 s, and forced from rest
by only the coupled wave radiation stresses. The \cd{NEMS} coupling time step
for the two model components is 1 s.                

\begin{table}[p]
\caption{Boers 1996 Wave conditions.  \\ \textit{All model configurations and
    results are pre-decisional and for official use only}.} 
\label{tab:boers}

\centering
\begin{tabular}{ c  c  c  }
\hline
\textbf{Wave condition} & \textbf{$H_s$ [m]} & \textbf{$T_p$ [s]}\\
\hline

1A & 0.157 & 2.05\\
1B & 0.206 & 2.03\\
1C & 0.103 & 3.33\\
\hline
\end{tabular}
\end{table}

\begin{figure}
    \centering
        \centering
    \hspace*{-12mm}
    \includegraphics[width=1.2\textwidth]{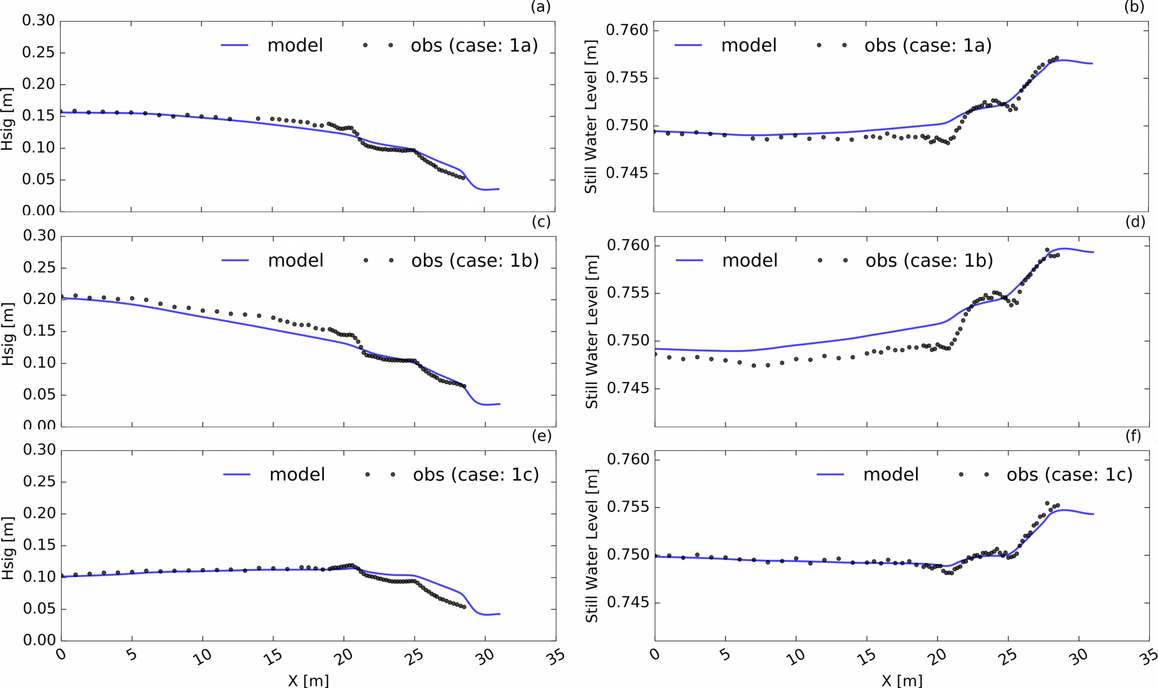}
    \caption{Significant wave height (a, c, e) and wave setup (b, d, f) for
    three different wave conditions which are summarized in Tab.
    \ref{tab:boers}. \\ \textit{All model configurations and
    results are pre-decisional and for official use only}.}
    \label{fig:boer_cases}
 \end{figure}


Fig. \ref{fig:boer_cases} shows that the modeled significant wave heights and
the wave setup produced by coupled \cd{ADCIRC-WW3} application are in good
agreement in terms of significant wave heights and the maximum water surface for
all three cases. For the more energetic wave case 1B, wave heights are somewhat
underestimated offshore of the outer bar. This overestimated dissipation appears to result in
the overestimation of the wave setup over this region leading up to the outer
bar. Conversely, in the milder-breaking swell case 1C we observe an
overestimation of wave heights inshore of the outer bar, presumably due to
insufficient depth-induced breaking. This results in a slight underestimation of
nearshore water levels. Nevertheless, these results show good overall skill of
our coupled \cd{ADCIRC-WW3} application in capturing the wave height and wave setup
evolution correctly.          


\subsection{Full scale inundation case}
\label{sec:hurr-ike}
\begin{figure}
    \centering
    \hspace*{-12mm}
    \includegraphics[width=0.8\textwidth]{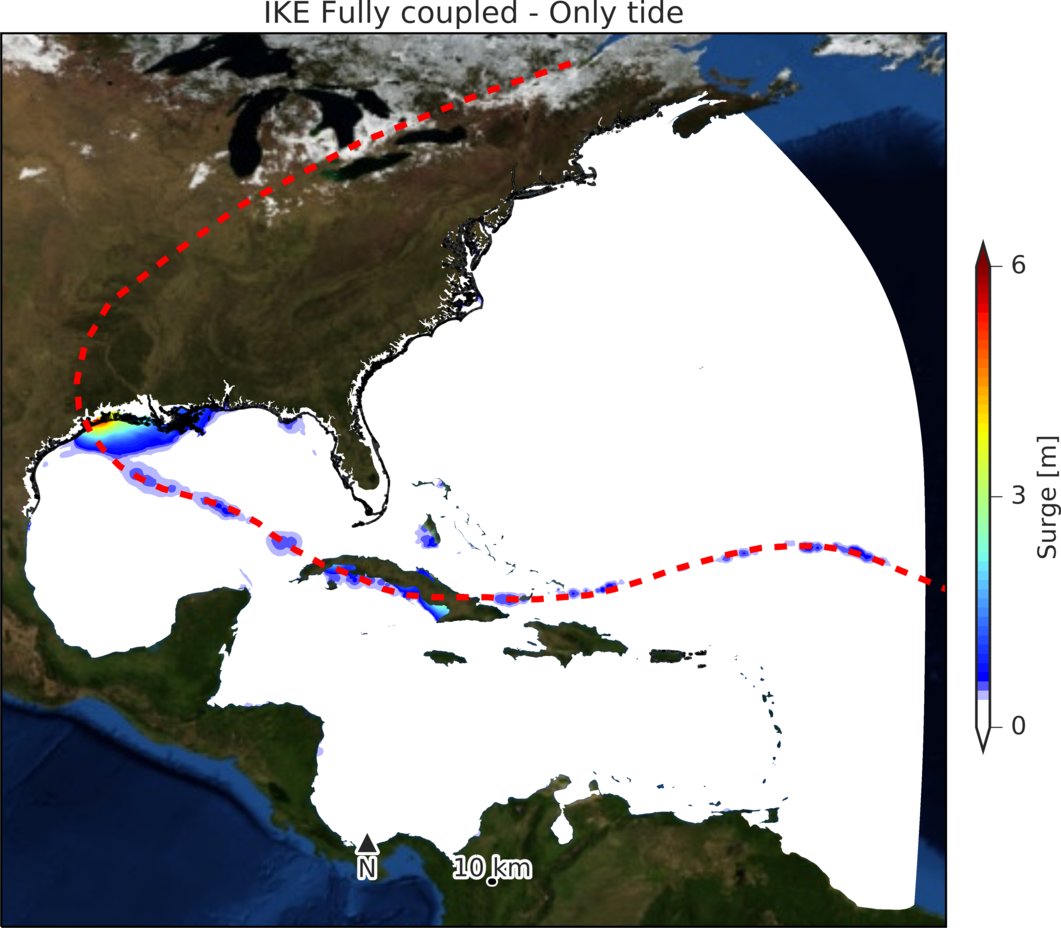}
    \caption{Surge level is computed by subtraction of tide elevation from maximum
total water level for the whole HSOFS mesh. Hurricane Ike’s best track is shown
by a red dashed-line. \\ \textit{All model configurations and
    results are pre-decisional and for official use only}.}
    \label{fig:hsofs-mesh_surge}
 \end{figure}

Here we present the results of the full scale inundation study for Hurricane
Ike, 2008. To simulate the combined waves and storm surge for Hurricane Ike, 
we utilized the unstructured triangular mesh
used by NOAA's operational Hurricane Surge On-demand Forecast System (HSOFS). We
applied 8 tidal constituents (M2, K1, O1, P1, Q1, N2, S2, K2) at
the model open boundaries (Figs. \ref{fig:hsofs-mesh_surge}).

\subsubsection{Atmospheric forcing}
\label{Sec:Atmosphericforcing}
The atmospheric forcing for this study is provided by NOAA's Hurricane Weather
Research and Forecasting (\cd{HWRF}) modeling system, coupled to the \cd{MPIPOM} ocean model,
and empowered by a movable multi-level nesting technology
\citep{zhang2016representing,hwrf2017}. The model grid is triple-nested using telescopic,
two-way interactive horizontal grid resolutions from synoptic with 
0.18\(^\circ\) resolution as the outer domain (spanning about 75\(^\circ\)
\(\times\) 75\(^\circ\)),  to moving storm nest with 0.06\(^\circ\) resolution
(10\(^\circ\) \(\times\) 10\(^\circ\)) and core of about 6\(^\circ\) \(\times\)
6\(^\circ\) with 0.02\(^\circ\) resolution. These nests follow the hurricane
best track, ensuring the highest resolution around the eye of a hurricane. In
this study, we have interpolated the hourly \cd{HWRF} model outputs from
multiple cycles initiated with analysis data and 9 forecast time steps. Every 6
hours, reanalysis data from the next cycle are smoothly ramped into the wind and
pressure fields. The atmospheric forcing has been validated against National
Data Buoy Center (NDBC) and satellite altimeter data. We extracted wind velocity
at 10 m height and surface pressure from the original \cd{GRIB2} output files
and saved them in \cd{NetCDF} format. The \cd{ATMesh} \cd{NUOPC} data cap reads
the meteorological forcing from \cd{NetCDF} file and provides it to \cd{ADCIRC}
and \cd{WW3} caps at every coupling time step. The \cd{HWRF} model was forced
with initial and boundary conditions provided by NOAA's Global Forecast System
(\cd{GFS}) with 0.5$^\circ$ spatial grid resolution.                      

\begin{figure}
    \centering
        \centering
    \hspace*{-12mm}    
   \includegraphics[width=1.2\textwidth]{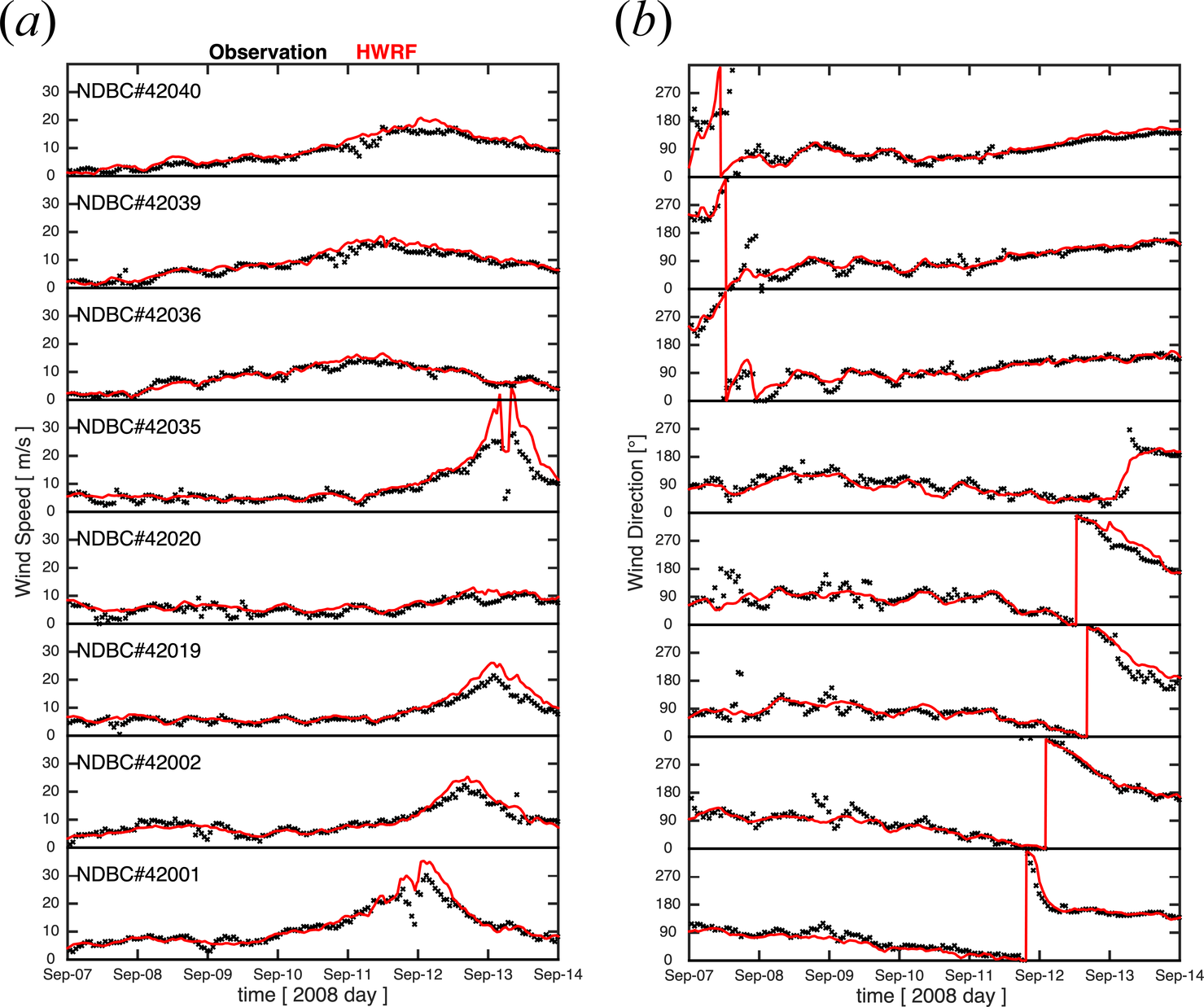}
   \caption{Atmospheric Model validation at NDBC buoy locations, Model (red)
   versus observation (black): (a) Wind speed (b) Wind direction. \\ 
   \textit{All model configurations and
    results are pre-decisional and for official use only}.}     
    \label{fig:HWRF}
 \end{figure}

Fig. \ref{fig:HWRF} gives an impression of the quality of the \cd{HWRF}
atmospheric forcing fields used to force the coupled \cd{ADCIRC-WW3} model, by
comparing 10 m wind speed and direction against NDBC buoy observations in the
Gulf of Mexico (see Fig. \ref{fig:DIF_CO_ST} for locations). We can see that the agreement is generally good, in particular
for wind directions. However, a tendency to overestimate the wind speeds at the
storm peak is found at the mid-Gulf NDBC buoys 42001 and 42002, and the shelf
buoy 42019. The high bias in the wind speed is particularly evident at landfall,
as seen at NDBC 42035, located on the shelf just offshore of Galveston. It would
be expected that these overestimated winds would lead to a degree of
overestimation of the locally-generated significant wave heights. It
is also worth mentioning that the NOAA buoy 42035 broke free during the storm
(See Blockedhttps://www.ndbc.noaa.gov/hurricanes/2008/ike/). Therefore, the
observation time series might not be at the same location of extraction 
coordinate from the models.

\subsubsection{\cd{ADCIRC} model component validation}
We validated the total water level results against two sets of water level
observations. The tidal gauge time-series were measured by the NOAA's Center for
Operational Oceanographic Products and Services (CO-OPS) and High Water Marks
(HWM) were measured and provided by United States Geographical Survey (USGS).

\begin{figure}
    \centering
    \hspace*{-12mm}
    \includegraphics[width=0.95\textwidth]{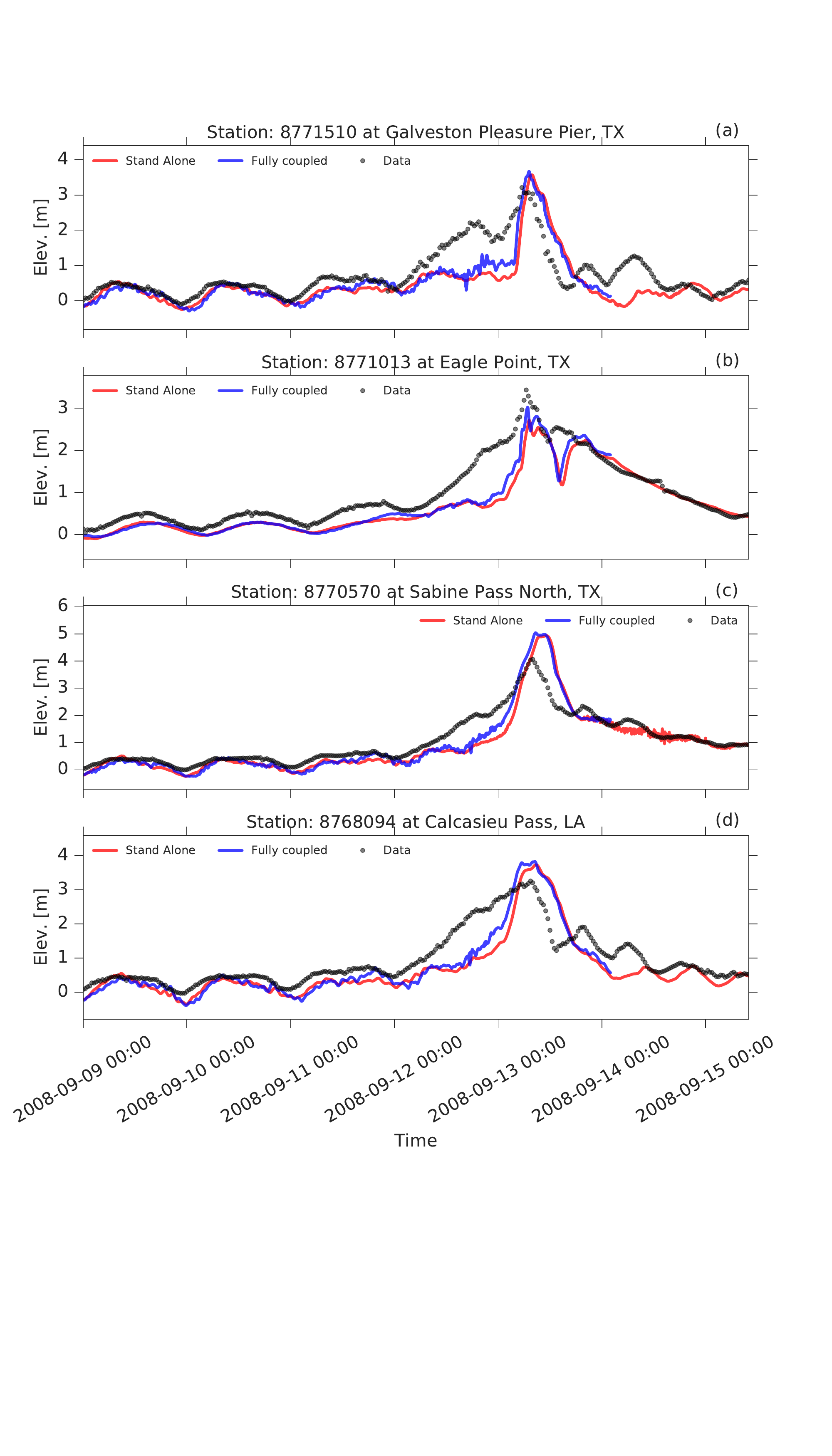}
    \caption{Time series of the total water level observations at the tidal gauges
(locations shown in Fig. \ref{fig:ts-map}). Black dots are the observations. Red line is the
tide only water level. Blue (GFS05d$\_$OC) and green (GFS05d$\_$OC$\_$Wav) are storm
induced total water level without and with wave forcing. Station names and ID
numbers are shown in each panel titles. \\ \textit{All model configurations and
    results are pre-decisional and for official use only}.}
    \label{fig:ts}
 \end{figure}

Comparison between the total water surface elevation from the coupled \cd{ADCIRC-WW3} application and tidal
gauges time series for four stations close to the actual hurricane track (location 
of the tidal gauges in Fig.~\ref{fig:ts-map} are presented in
Fig.~\ref{fig:ts}a-d).
The modeled water surface elevation agrees well with the observation in terms of the 
level and timing of the peak of the storm inundation. 

\begin{figure}
    \centering
    \hspace*{-12mm}
    \includegraphics[width=0.95\textwidth]{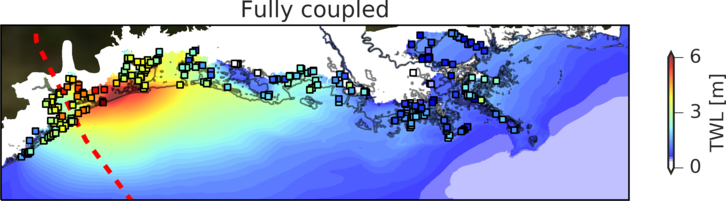}
    \caption{High water marks observations for Hurricane Ike, 2008. The contour
    plot is the total water level for GFS05d$\_$OC$\_$DA$\_$Wav case. \\ \textit{All model configurations and
    results are pre-decisional and for official use only}. }
    \label{fig:hwmmap}
 \end{figure}

High Water Marks are an important source of observations for validation and
enhancement of the storm surge and flood inundation studies. After significant
flooding due to a land-falling hurricane, a rapid high water mark (HWM) data
collection by USGS takes place to document the event and to help improving
future disaster preparedness activities. Comparison of the total water level from the 
coupled \cd{ADCIRC-WW3} application and HWMs observation reveal that both are 
in general agreement in particular around the hurricane track in the land-fall 
region (Fig. \ref{fig:hwmmap}). 

\begin{figure}
    \centering
    \hspace*{-12mm}
    \includegraphics[width=0.95\textwidth]{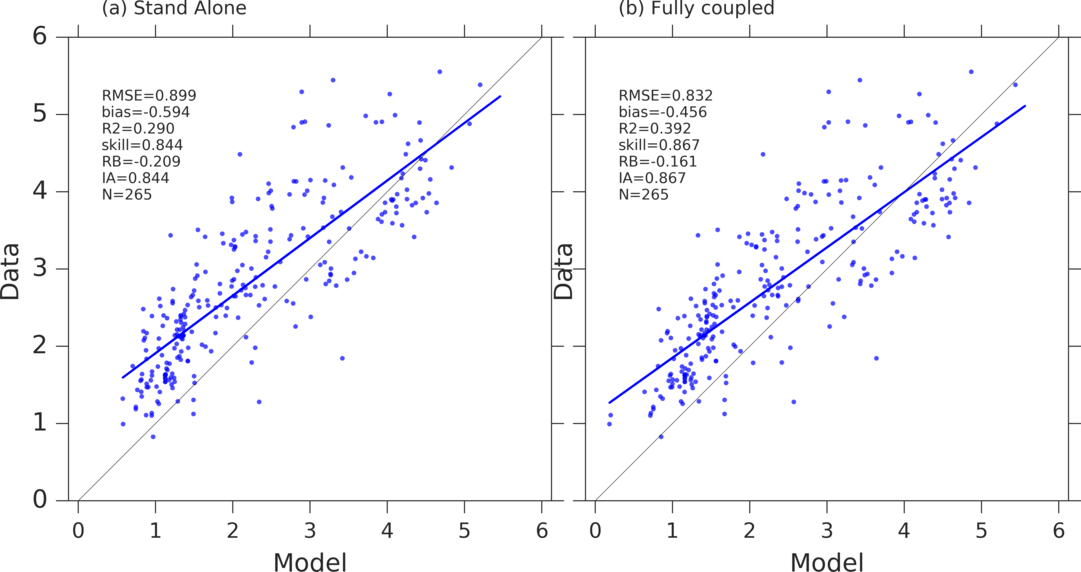}
    \caption{Statistical comparisons of High water marks observation and model
    results. Location of the high water mark observations are shown in Fig.
\ref{fig:hwmmap}. \\ \textit{All model configurations and
    results are pre-decisional and for official use only}.}
    \label{fig:hwm_scatter}
 \end{figure}

We also compared scatter plots and a number of statistical metrics of the total
water levels and HWM data (Fig. \ref{fig:hwm_scatter}). Both Fully coupled and 
Stand alone results show
underestimation of the model in comparison to observations. However, upon 
inclusion of the wave forcing, an improvement in the error statistics is found, such 
that a reduction in the relative bias (RB) from -0.594~m to -0.392~m, and a reduction 
in root mean square error (RMSE) from 0.899~m to 0.832~m were presented.

\subsubsection{\cd{WW3} component verification}
The coupled \cd{ADCIRC-WW3} coupled application performance with respect to the
wave modeling aspects was firstly assessed in the offshore and over the shelf of the Gulf of Mexico using 
NDBC monitoring buoys, and secondly at the location of landfall by 
using rapid-deployment pressure sensors, placed by \cite{kennedy2011origin}.

\begin{figure}
   \centering
       \centering
    \hspace*{-12mm}
   \includegraphics[width=1.2\textwidth]{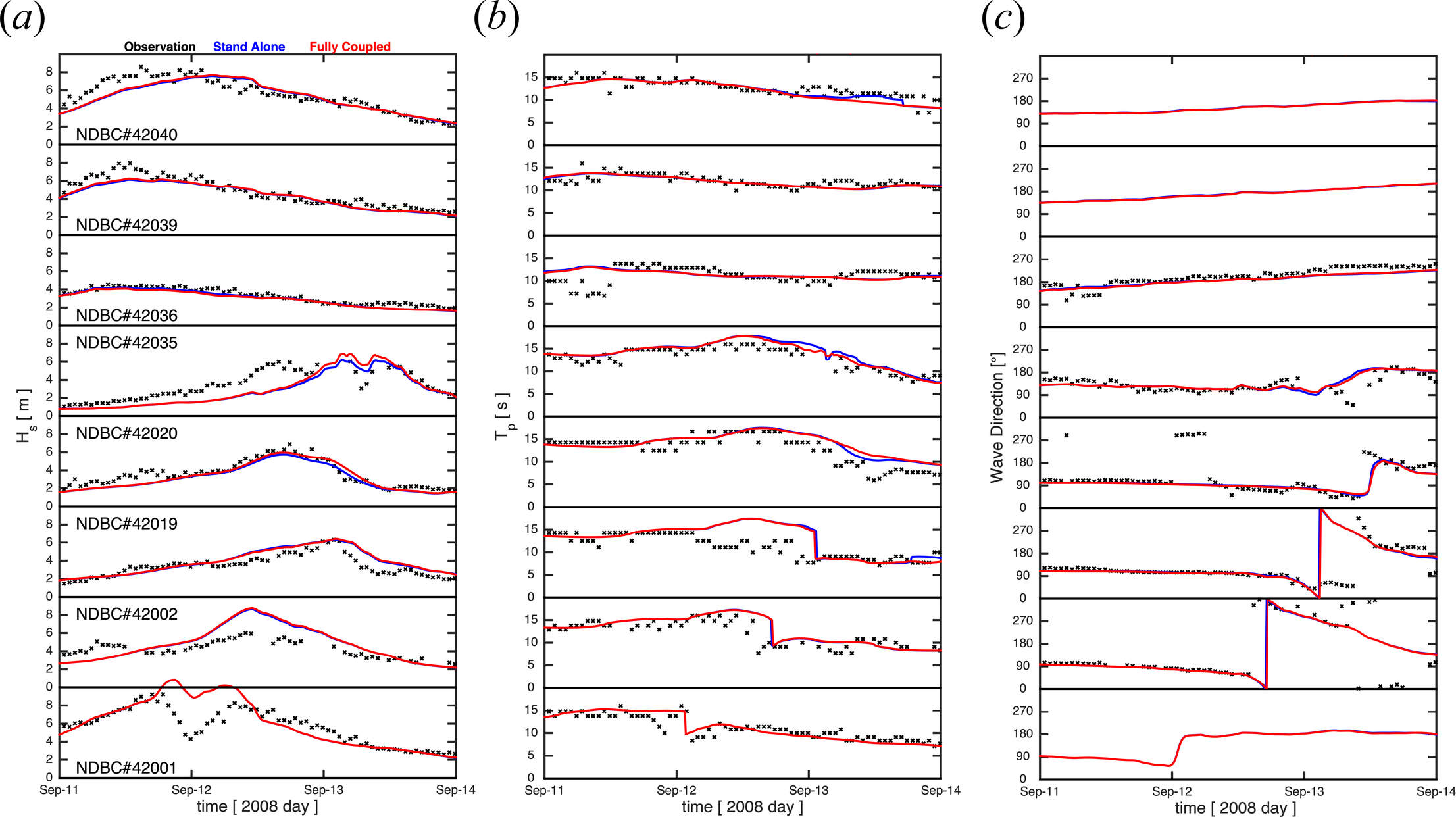}
   \caption{Model validation at NDBC buoy locations, forced by HWRF atmospheric
   model, extracted from stand-alone WW3 simulations (blue), fully coupled
   Wave-Surge simulation (red) versus observation (black): (a) Significant wave
   height ($H_s$); (b) peak period ($T_p$) and (c) mean wave direction. All
   model configurations and results are pre-decisional and for official use
   only. \\ \textit{All model configurations and  results are pre-decisional and
   for official use only}.}   
    \label{fig:WW3NDBC}
 \end{figure}

Considering the NDBC monitoring buoys, four groups can be distinguished. At mid-Gulf 
(NDBC 42001 and 42002), we find some overestimation of significant wave heights 
during the passing of the hurricane in response to the somewhat overestimated 
wind speeds seen above (Fig. \ref{fig:WW3NDBC}a). Similarly, the peak wave periods 
are overestimated, reflecting excessive wind-sea growth (Fig. \ref{fig:WW3NDBC}b). 
However, wave direction is reproduced well (Fig. \ref{fig:WW3NDBC}c). Interestingly, 
in the far-field, towards the eastern half of the Gulf (NDBC 42036, 42039 42040), 
significant wave heights tended to be underestimated during the passing of the 
storm, while the peak period and direction were captured well. Moving onto the 
shelf in the region of landfall (NDBC 42019 and 42020) the agreement of the model 
results with the observations improved in general, although there is still 
overestimation in the significant wave height and peak period. Finally, at 
NDBC 42035, just offshore of the landfall location at Galveston, all wave model 
parameters are in good agreement with the observations at the storm peak. However, 
the significant wave height is underestimated at this shallow water location 
just ahead of the storm peak which could be related to the omission of the
forerunner effect in the coupled surge model component. Regarding the differences between the 
coupled and stand-alone versions of the wave model, these can only be seen, 
as expected, at the two shallower stations NDBC 42020 and 42035. As expected, 
the increased water levels seen above increase the modeled 
significant wave height at these stations.

It should be noted that our concentration here is to ascertain general
performance of the coupled model. In terms of forerunner, we reference our reader to
\cite{kennedy2011origin}. We diagnose the forerunner surge as being generated 
by Ekman setup on the wide and shallow shelf. The longer forerunner time 
scale additionally served to increase water levels significantly in 
narrow-entranced coastal bays. 


\begin{figure}
    \centering
        \centering
    \hspace*{-12mm}
    \includegraphics[width=1.2\textwidth]{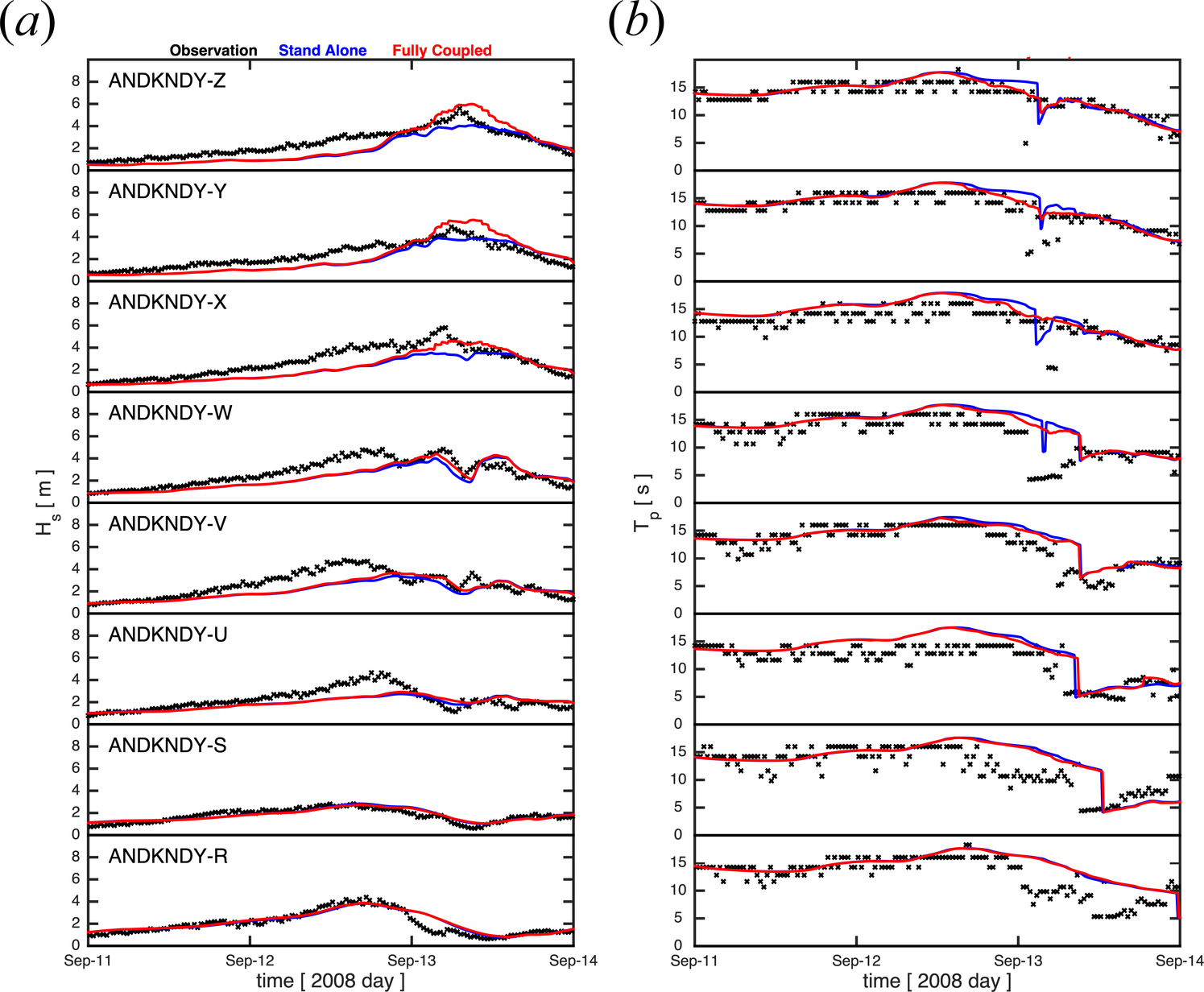}
  \caption{Model validation at quick deployed gauges, forced by HWRF atmospheric
  model, extracted from stand-alone WW3 simulations (blue), fully coupled
  Wave-Surge simulation (red) versus observation (black): (a) Significant wave
  height ($H_s$); and (b) peak period ($T_p$). All model configurations and
  results are pre-decisional and for official use only. \\ \textit{All model
  configurations and      results are pre-decisional and for official use  
  only}.}
    \label{fig:WW3AND}
 \end{figure}

Moving more towards shoreline, model results at the pressure sensors deployed 
in the surf zone by \cite{kennedy2011origin} show a significant sensitivity
 to the inclusion of variable water levels from the coupled ADCIRC model (Fig.
 \ref{fig:WW3AND}). Referring to Fig. \ref{fig:DIF_CO_ST}b for the station
 locations, the most significant of these are the stations ANDKNDY-X, Y and Z
 located under the eastern half of the landfalling hurricane. For these three
 stations, we can see a large influence of the added surge level on the
 significant wave height at the storm peak, in all cases improving the agreement
 with observations. 
 We note that, as discussed above, the stations U to Z also might 
 relate to underestimation of the water surface elevation during forerunner.
 Stations S and R, located to the
 south of the landfall location received mostly offshore winds and a water level
 set-down. The significant wave heights at these stations are reproduced well.          

\begin{figure}
    \centering
    \centering
    \hspace*{-12mm}
   \includegraphics[width=1.2\textwidth]{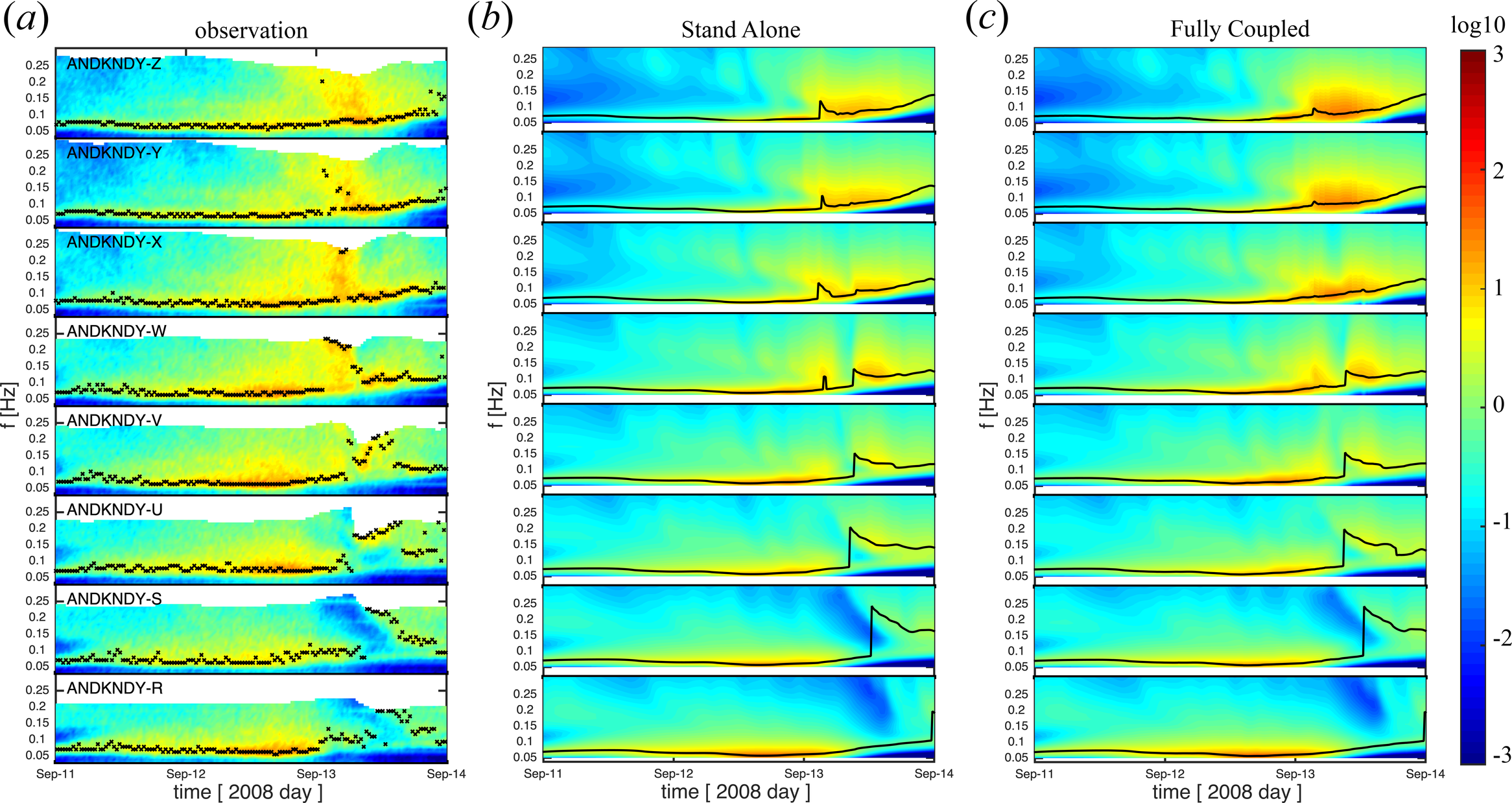}
\caption{Frequncy spectrum at nearshore quick deployed gauges comparing
observation (a) versus stand-alone WW3 simulations (b) and fully coupled
Wave-Surge simulation (c). The time series of peak frequency is shown in each
subplot.  All model configurations and results are pre-decisional and for
official use only. \\ \textit{All model configurations and  results are
pre-decisional and for official use only}.}    
    \label{fig:SpecAND}
 \end{figure}

Fig. \ref{fig:SpecAND} provide a more in-depth view of the results at 
the \cite{kennedy2011origin} nearshore stations by considering their spectrograms. 
This figure shows a number of important features of the nearshore wave field 
transformation under both the coupled and stand-alone models. Up to September 
13, the simulated variance density at the northerly stations ANDKNDY-X, Y and Z 
is underestimated by both the stand-alone (Fig. \ref{fig:SpecAND}b) and coupled 
(Fig. \ref{fig:SpecAND}c) models in connection with non-resolved forerunner
effects.
After September 13, the variance density abruptly increases, as the main storm peak 
arrives with landfall. At this time, the coupled model (panel c) shows greater 
levels of variance density than the stand-alone model (panel b), which agrees 
better with the observations. A final important feature of the nearshore spectra 
is the frequency upshift of the spectral peak, due to the nonlinear 
three-wave interactions. This strong upshift at stations V, W, X, Y, Z is not 
seen as intensely in the model results, indicating an underestimation 
of the magnitude of this process by the LTA nonlinear interaction source term. 
Interestingly, the stand-alone model captured this upshift process somewhat 
better than the coupled model, since the depth underestimation in the former 
fortuitously enhances the computed nonlinear interaction (see stations X, Y, Z).

\section{Summary and Conclusions}
\label{sec:Summary}

We developed a flexible coupling application for coastal inundation studies in the framework of
\cd{NEMS} using \cd{NUOPC/ESMF} infrastructure. This application includes
\cd{NUOPC} model interfaces (or caps) for \cd{ADCIRC} and \cd{WW3}, and also a
data cap to read and provide atmospheric forcing fields from \cd{HWRF} model
results. The application is examined using a standard laboratory flume case for 
set-up and wave dissipation, and validated for Hurricane Ike, 2008 on NOAA's
HSOFS mesh, a 1.8~M node triangular mesh with a nominal resolution of $\sim$500
m. The model skills and improvement due to wave effects on the final inundation were examined and discussed using
time series from tide gauges and high water marks observations.

In general, it can be concluded that the coupling improves the performance of
the \cd{ADCIRC} and \cd{WW3} model components. The \cite{boer1996Surfzone} case
shows that the coupling between these two models behaves correctly under laboratory conditions. In the Ike field case, the simulated water
levels of \cd{ADCIRC} showed generally better agreement with observations upon
inclusion of the coupled wave effects from \cd{WW3}. Accounting for the
forerunner effect is expected to improve the results further. Conversely, the 
coastal wave field simulated by \cd{WW3} improves significantly
upon coupling with \cd{ADCIRC}, in particular in the surf zone and in inundated
regions.

As outlined in the Introduction, the first application of the \cd{NUOPC} model
combination (or App) described here is the Named Storm Event Model (NSEM), a high-resolution hindcast model
which is being developed to meet the requirements of the COASTAL Act (2012). However, we
anticipate that these flexible and generic NUOPC coupling caps for ADCIRC and WW3 will
enable future development and seamless inclusion of various additional model components for
forecasting applications. For example, the current cap development provides the possibility to
perform three-way dynamical coupling of \cd{ADCIRC}, \cd{WW3} and atmospheric
prediction models.
In the hindcast application described here, we limited the atmospheric model, run offline, to a
one-way forcing to \cd{ADCIRC-WW3} via a data cap. However, the \cd{ADCIRC} and
\cd{WW3} caps are capable of dynamically importing atmospheric model forcing,
and exporting current velocity, water surface elevation (accounting for inundated zones which alter the surface roughness), and
enhanced sea surface roughness due to the presence of waves (Charnock parameter) back to the
connected atmospheric model component.

Further future coastal applications of the presented flexible \cd{NUOPC}
modeling framework are incorporating processes such as river and inland flooding and sea ice. As part of the Named
Storm Event Model, work is currently under way to incorporate river flow into this coupled
system by including a \cd{NUOPC} cap for \cd{WRFHydro}. This
extension will allow the modeling of complex coastal flooding events such as Hurricane Harvey (2017), which featured
extreme rainfall and inland flooding alongside the surge and wave forcing from the ocean. Sea
ice has significant and complex impacts on coastal surge
\citep{westerink2018coupled} as well as dissipative and dispersive effects on the coastal 
wave field \citep{rogers2018forecasting}. The
flexible coastal modeling system presented here can be extended to include these coupled processes via
an existing \cd{NUOPC} cap for the sea ice model \cd{CICE}
\citep{campbell2013coupling}.

\section*{Acknowledgements}
This work has been supported by The Consumer Option for an Alternative System to Allocate Losses
(COASTAL) Act Program and Water Initiative project within the National
Oceanic and Atmospheric Administration (NOAA). The WaveWatchIII development has been supported by the US. 
Army Corps of Engineers (USACE). The authors acknowledge Dr. Andrew Kennedy for providing the
in-situ data. The results presented here are preliminary and for presenting the
incremental developments of coupled application which is under development to
serve above mentioned projects. The authors also would like to acknowledge Dr.
Chris Massey for his constructive communications.

\bibliographystyle{spbasic}

\nolinenumbers
\newpage
\section*{Appendix A: Metrics for the evaluation of data-model agreement}
\label{sec:appendix}

In order to assess model performance for water levels, root mean square error
(RMSE), BIAS, relative BIAS (RB), Correlation (Cor), Index of Agreement (IA) and
peak error (Peak) were used.

The RMSE is given by
\begin{equation}
 RMSE = \sqrt{\frac{1}{N} \sum\limits_{i=1}^N (M_i - O_i)^2}
\end{equation}
where $M_i$ is the modeled data, $O_i$ is the measured data and \textit{N} is the total number of data.

BIAS shows the systematic deviation from the observations and is given by
\begin{equation}
 BIAS = \sum\limits_{i=1}^N (M_i - O_i)
\end{equation}

Relative BIAS (RB) shows relative systematic deviation from the observations
and is given by
\begin{equation}
 RB = \frac{\sum\limits_{i=1}^N (M_i - O_i)}{N \langle O_i \rangle } 
\end{equation}

Peak error is calculated by
\begin{equation}
 PEAK = \max{O} - \max{M} 
\end{equation}

The Index of Agreement (IA) is formulated as
\begin{equation}
 IA = 1 - \frac{ \sum\limits_{i=1}^N (M_i - O_i)^2}
 {\sum\limits_{i=1}^N \left(\left(\vert M_i - \langle O \rangle \vert + \vert O_i - \langle O \rangle \vert  \right) ^2 \right)}
\end{equation}
where brackets, $\langle \cdot \rangle$, denote time averaging. $IA=1$ shows perfect agreement and $IA=0$ means complete disagreement.

The Pearson correlation (Cor) coefficient is calculated by
\begin{equation}
 Cor =\frac{ \sum\limits_{i=1}^N (O_i -  \langle O \rangle) (M_i -  \langle M \rangle) } 
  {\sqrt{ \sum\limits_{i=1}^N (O_i -  \langle O \rangle)^2  \sum\limits_{i=1}^N
 (M_i -  \langle M \rangle)^2 }}
\end{equation}
It has a value between $+1$ and $−1$, where $1$ is total positive linear
correlation, $0$ is no linear correlation, and $−1$ is total negative linear correlation 

\end{document}